# Scanning tunneling microscopic evidence of interface enhanced high-$T_c$ superconductivity in Pb islands grown on $SrTiO_3$

Haigen Sun[1], Zhibin Shao[1], Zongyuan Zhang[1], Yan Cao[1], Shaojian Li[1], Xin Zhang[1], Qi Bian[1], Habakubaho Gedeon[1], Hui Yuan[1], Jianfeng Zhang[2], Kai Liu[2], Zhong-Yi Lu[2], Tao Xiang[3,4], Qi-Kun Xue[4,5] and Minghu Pan[1]

[1]School of Physics, Huazhong University of Science and Technology, Wuhan 430074, China.
[2]Department of Physics，Renmin University of China，Beijing 100872，China
[3]Institute of Physics, Chinese Academy of Sciences, Beijing 100190, China
[4]Collaborative Innovation Center of Quantum Matter, Beijing 100084, China
[5] Department of Physics, Tsinghua University, Beijing 100084, China

Haigen Sun, Zhibin Shao, Zongyuan Zhang contributed equally to this work.
Corresponding author. E-mail: qkxue@mail.tsinghua.edu.cn (Q.-K. X.); minghupan@hust.edu.cn (M. H. P.).

**The discovery of the interface enhanced superconductivity in the single layer film of FeSe epitaxially grown on $SrTiO_3$ substrates has triggered a flurry of activity in the field of superconductivity. It raised the hope to find more conventional high-Tc superconductors which are purely driven by the electron-phonon interaction at ambient pressure. Here we report the experimental evidence from the measurement of scanning tunneling spectroscopy for the interface enhanced high-Tc superconductivity in the Pb thin film islands grown on $SrTiO_3$ substrates. The superconducting energy gap of the Pb film is found to depend on both the thickness and the volume of the islands. The largest superconducting energy gap is found to be about 10 meV, which is 7 times larger than that in the bulk Pb. The corresponding superconducting transition temperature, estimated by fitting the temperature dependence of the gap values using the BCS formula, is found to be 47 K, again 7 times higher than that of the bulk Pb.**

Searching of high-temperature superconductors remains one of the most important challenges in condensed matter physics. A conventional wisdom, motivated by the unconventional superconductivity discovered in cuprate and iron-based superconductors [1,2], is to push a material close to an antiferromagnetic or other kind of instabilities either by chemical doping or by applying high pressure so that strong quantum fluctuations can drive it into a high-$T_c$ superconductor. For conventional superconductors whose superconducting pairing is glued by electron-phonon coupling, high-Tc superconductivity can be also achieved by metalizing sigma-bonding electrons [3]. A typical example is $MgB_2$ ($T_c \sim 39$ K) where a two-dimensional like sigma-bonding band of the in-plane B 2p electrons is lifted above the Fermi level by an attractive interaction generated by $Mg^{2+}$ cations. Recently, the discovery of superconductivity in the single layer film of FeSe epitaxially grown on $SrTiO_3$ (001) substrates has opened a new route to find high-Tc superconductivity [4]. The superconducting gap ($\Delta \sim 15$-$20$ meV) and the corresponding superconducting transition temperature ($\sim 65$ K) of FeSe/$SrTiO_3$ are about one order of magnitude higher than those in the bulk FeSe ($\Delta \sim 2.2$ meV, Tc $\sim 8$ K). The large enhancement of superconductivity has also been found in the FeSe films grown on $SrTiO_3$ (110) [5] or $BaTiO_3$ substrates [6], or in the partial Te substituted FeSe film, i.e. $FeTe_{1-x}Se_x$ [7], grown on $SrTiO_3$ (001) [8]. Besides, a similar superconducting enhancement induced by $SrTiO_3$ (001) substrate was recently observed in conventional superconductor Sn islands [9]. It is believed that both the interface induced charge transfer [6,8,10,11] and the interface-enhanced electron-phonon coupling [11,12] play an important role in this enhancement of superconductivity.

During the past two decades, lead (Pb), an elemental superconductor, grown on Si (111) substrate has been extensively studied. Various novel phenomena, such as the oscillatory transition temperature with the thickness [13,14] and the island-size-dependent superconductivity [15], were reported. However, a significant interface-enhanced superconductivity has not been observed in Pb films or islands on Si substrates [13,15,16].

In this communication, we report the results of scanning tunneling spectroscopy measurements for Pb thin films grown on $SrTiO_3$ substrates. On each film, Pb atoms form many hexagonally ordered and dense packed islands whose sizes vary from 286 nm³ to 4945 nm³. The superconducting states are observed when the volume is above 1200 nm3. The tunneling spectrum, measured in a relative large Pb island, exhibits a U-shaped superconducting gap of 7.1 meV at 0.4 K. The energy gap of this Pb island eventually vanishes at 32 K. The gap increases with decreasing volume of Pb islands. The largest superconducting energy gap we found from all the islands measured is about 10 meV, which is more than 7 times higher than the corresponding value (1.4 meV) in the bulk Pb. This gap decreases with increasing temperature, but remains finite even at the the highest temperature (33 K) our STM measurement can be done. By fitting the temperature dependence of the gap values using the BCS gap formula, we find that the transition temperature can be as high as 47 K.

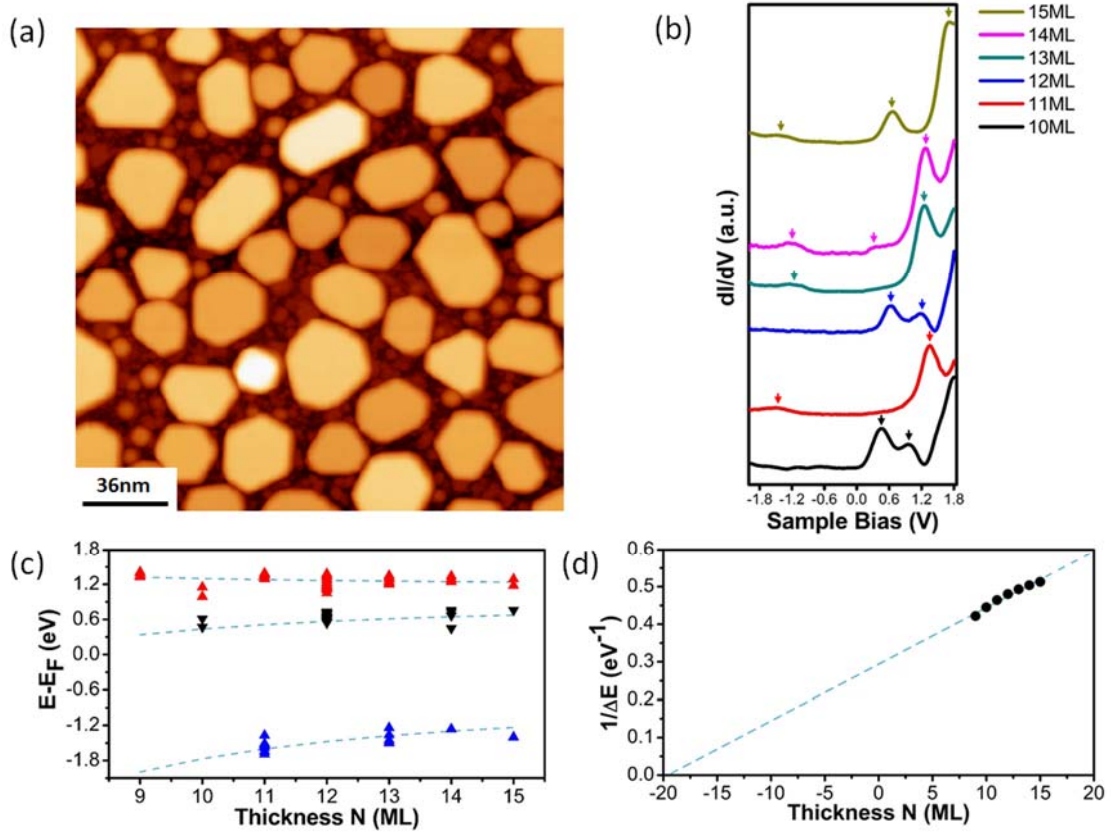

**Figure 1 | Topographic image and quantum well states.** (a) A topographic image of Pb islands grown on a SrTiO₃ substrate taken at 4.2K. The hexagonally ordered shape of the islands is a manifestation of good crystallization. (b) Differential conductance, dI/dV, measured at the center of six Pb islands with different layer thicknesses. The arrows correspond the peaks of quantum well states. The bias voltage and tunneling current were set to 0.5V and 0.1nA. (c) The thickness dependence of the peak energies

of quantum well states. The dashed lines are the fitting curves obtained using the quantized energy level formula of the one-dimensional infinite-well model. (d) The inverse of the energy separation of adjacent quantum well states near the Fermi level as a function of thickness. The dashed line is a linear fit to the data.

**Topographic STM image and the distribution of quantum well states**

Figure 1(a) shows a typical topographic image of Pb islands. In the synthesis of the film [17], a wetting layer of Pb is first grown on the surface of SrTiO$_3$ to release the strain induced by the lattice mismatch. The Pb islands are then grown above the wetting layer. In comparison with the morphology of Pb islands on Si (111) [18, 19], the wetting layer between the islands and the SrTiO$_3$ substrate is more disordered, which suppresses the diffusion between different islands, preventing the formation of large islands by merging together two or more nanoscale-sized islands. Thus the Pb islands formed on the SrTiO$_3$ substrate distribute densely. The hexagonal order revealed by these Pb islands indicates that the film is well crystallized. The sizes of Pb islands vary from 9 to 23 monolayers in thickness and 286-4945 nm$^3$ in volume.

In order to check whether quantum well states are formed on the Pb islands, we measured the tunneling spectra ranging from -2V to 1.85V at the center of six islands with different layer thicknesses. As shown in Fig. 1(b), well-defined quantum well state peaks, which are formed by multiple reflections of electrons between the surface and the interface of the island, were observed. The energies of the quantum well states can be read out from the peak positions of the spectral function. Figure 1(c) shows the energy of the quantum well state as a function of the island thickness N. We find that this energy scale can be described by the quantized energy level formula of the one-dimensional infinite-well model [20].

Furthermore, we find that the inverse of the energy difference $\Delta E$ between the two neighboring quantum well states near the Fermi level [the data represented by the black and blue triangles in Fig. 1(c)] scales linearly with the thickness N, i.e. $1/\Delta E \sim N$ [Fig. 1(d)]. The extrapolated $1/\Delta E$ crosses the thickness axis at a negative value, which implies that the effective width for confining electrons is larger than the thickness of a

Pb island. For the Pb islands grown on a Si(111) substrate, the extrapolated $1/\Delta E$ crosses the thickness axis at -3 ML [21, 22], which suggests that the effective thickness of the wetting layer is 2. However, for Pb islands grown on $SrTiO_3$, the extrapolated intersection is about -20 ML, exhibiting a huge increase in the effective width of the quantum well. This huge increase might be due to the charge transfer from the interface on which a 2D electron gas induced by high temperature annealing is expected to exist [10, 23, 24].

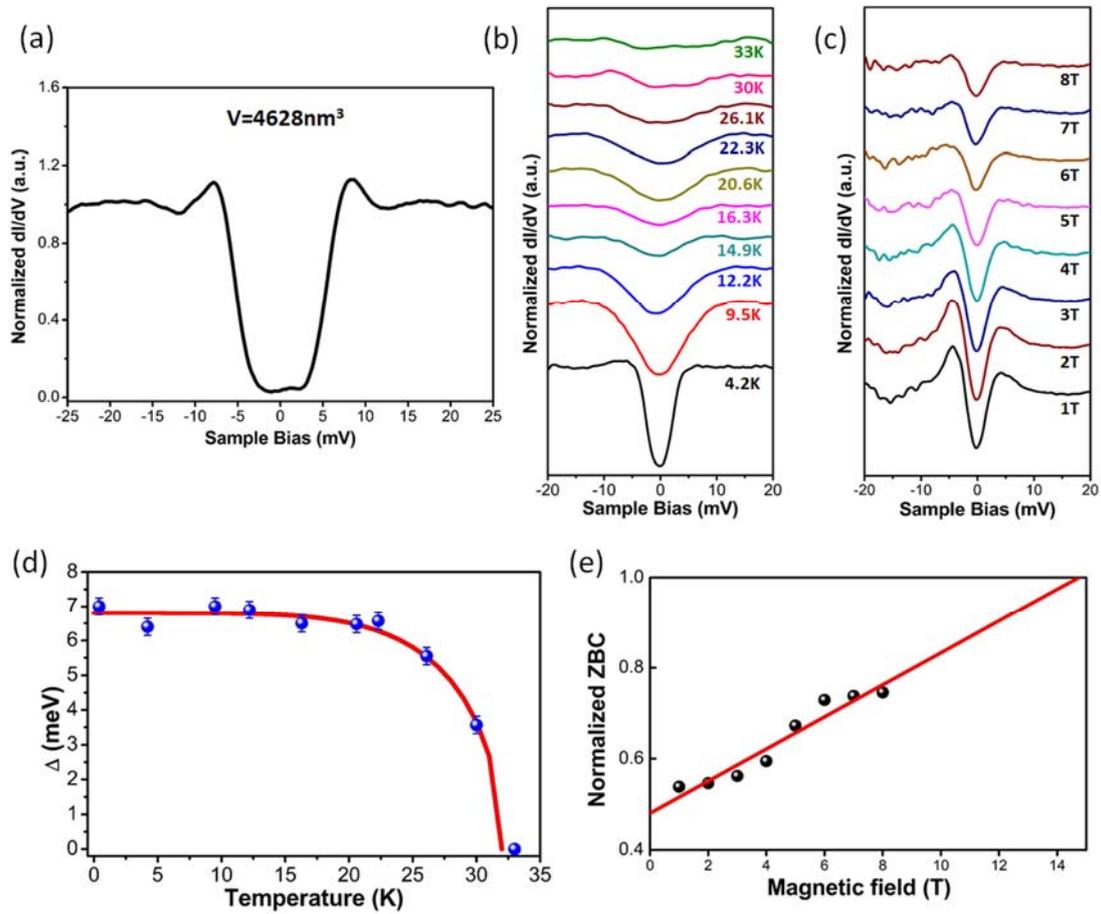

**Figure 2 | Temperature and magnetic field dependences of the tunneling spectra.** (a) dI/dV measured at the center of a Pb island with the volume V=4628 $nm^3$ at 0.4K, which shows a U-shaped spectrum with a superconducting gap about 7.1 meV. (b) Variances of dI/dV with temperatures. The spectra measured at different temperatures are shifted vertically for clarity. (c) dI/dV measured under several different magnetic fields at 4.2 K. (d) Temperature dependence of the superconducting gap extracted from (b). The red solid curve shows the fitting by the BCS gap function (see the text). (e)

Magnetic field dependence of the normalized zero bias conductance extracted from (c). The red line is a linear fit to the data. All dI/dV tunneling spectra are measured with a bias voltage of 30 mV and a tunneling current of 200 pA. The bias modulation is set at 0.5mV. dI/dV is normalized by dividing the normal-state conductance which is obtained by a cubic fit to the high-bias data with |V|> 10 mV.

**Temperature and magnetic field dependences of STS spectra**

To explore the superconductivity, high resolution spectra of differential conductance dI/dV were measured first at the center of a large Pb island with volume V=4628 nm$^3$. Figure 2(a) shows the normalized dI/dV spectrum taken at 0.4 K, which displays a clear signature of superconductivity with a U-shaped superconducting energy gap. From the half distance between the two conductance peaks, we find that the superconducting energy gap $\Delta$ is about 7.1 meV, which is about 5 times larger than the corresponding gap in the bulk Pb (1.40meV) [25]. This large enhancement in the energy gap has not been observed in Pb islands grown on Si(111) [15, 19, 26]. In Ref. [15], it was reported that a Pb island grown on Si(111) with a similar volume (V=4873 nm$^3$) just has a superconducting energy gap of 1.25 meV. These results suggest that the SrTiO$_3$ substrate plays a crucial role in the enhancement of superconductivity.

Figure 2(b) shows the temperature evolution of dI/dV from 4.2 K to 33 K. With increasing temperature, the conductance dip at the zero bias is reduced and eventually vanishes at about 33 K. The gap value obtained by fitting the experimental data with the Dynes function [27] as a function of temperature is shown in Fig. 2(d). Details of the fitting are presented in Fig. S1 of the Supplemental Information. By fitting the gap values using the BCS gap formula [28], we find that the zero temperature energy gap $\Delta(0)$ and the critical temperature $T_c$ are equal to 6.8 meV and 32 K, respectively. The BCS ratio, $2\Delta(0)/k_BT_c = 4.93$, is higher than that in a Pb island ($2\Delta(0)/k_BT_c = 4.5$) with almost the same volume grown on Si (111) substratec [15] or in the Pb Bulk [25]. This implies that the electron-phonon coupling [15] is stronger in the Pb islands grown on SrTiO$_3$ according to the Eliashberg theory [25, 29].

Figure 2(c) shows how the spectra vary with an external magnetic field applied perpendicular to the sample surface up to 8T at 4.2 K. As expected, the superconducting gap feature is suppressed by the applied field. The upper critical field is apparently higher than 8 T, which is already two orders of magnitude higher than the upper critical field in the bulk Pb (0.08 T) [30]. The normalized zero bias conductance extracted from Fig. 2(c) is depicted in Fig. 2(e).

The huge enhancement in the upper critical field is likely to be caused by the size effect, because a vortex can survive only when the in-plane dimension of Pb islands is above a critical length scale. Below the critical length scale, the applied field is less screened and the critical field is strongly enhanced [31, 32]. The effective diameter of the Pb island we measured here (20nm) is about three times smaller than the critical length for the vortex formation (60 nm) [19, 26]. Similar enhancement was reported in the Pb/Si(111) system [19] as well as in the superconducting Pb nano-particles [31].

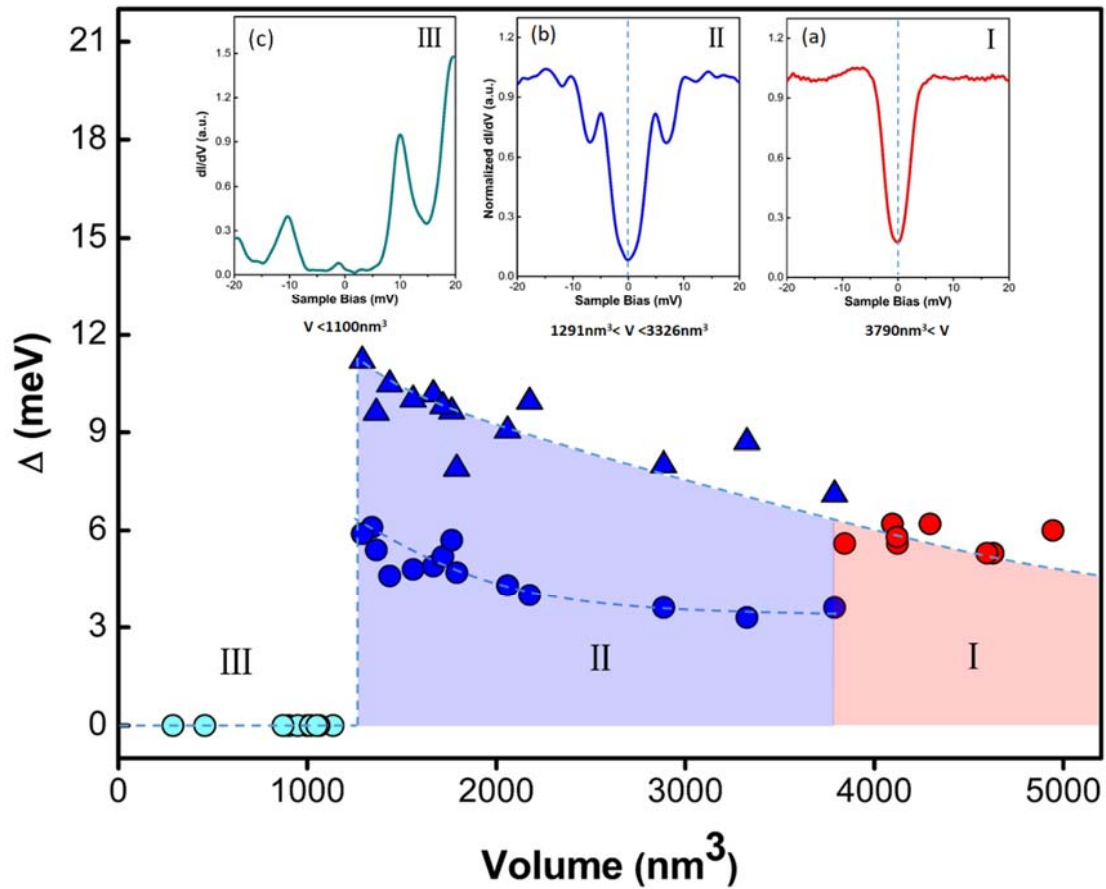

**Figure 3 | Volume-dependence of the superconducting energy gap.** The energy gap is divided into three distinct regions marked by the roman numerals and by different colors. The dash lines are provided to guide the eye. (a)-(c) Three representative

normalized dI/dV spectra measured in the corresponding regions. The magnitudes of double gaps in Regions II are depicted by blue filled triangles and circles, respectively.

**The volume dependence of STS spectra and superconducting gap**

In order to investigate how the superconducting energy gap changes with the island size, we measured the tunneling spectroscopy at 4.2K for all the islands shown in Fig. 1(a). The volumes of Pb islands vary from 286nm$^3$ to 4945 nm$^3$. From the line shape of the spectra, we find that the islands can be divided into three regions. In the region where the volume is larger than 3790nm$^3$ (Region I), the spectrum exhibits a single gap feature with two conductance peaks at gap edges smeared out by thermal effect (Fig. 3(a)). The conductance peak, as shown in Fig. 2(a), becomes more pronounced when the sample is cooled down to 0.4K. On the other hand, when the volume is less than about 1100 nm$^3$ (Region III), the superconducting gap vanishes. The spectra do not change much when the temperature varies from 4.2 K to 33 K. The two peaks at ±10 meV as shown in Fig. 3(c) are not the superconducting coherence peaks. They may result from strong phonon excitations at that energy scale.

When the volume is between 1291 nm$^3$ and 3326 nm$^3$ (Region II), a two-gap structure is observed in the tunneling spectra. The spectral function inside those two gaps varies strongly with temperatures (Supplementary Information Fig. S4), suggesting both are superconducting gaps. The inner two coherence peaks, which correspond to the smaller gap, emerge symmetrically at the positive and negative bias sides.

The main panel of Fig. 3 shows how the superconducting gaps vary with the volume of Pb islands in Regions I and II. The gap values are determined by the peak energies (the detail can be found from Fig. S2 in Supplementary Information). The larger gap increases with reduced volume from Region I to Region II. Region I is likely to be a two-gap system as well. But the inner gap is not observed in this region, probably due to the limitation of resolution.

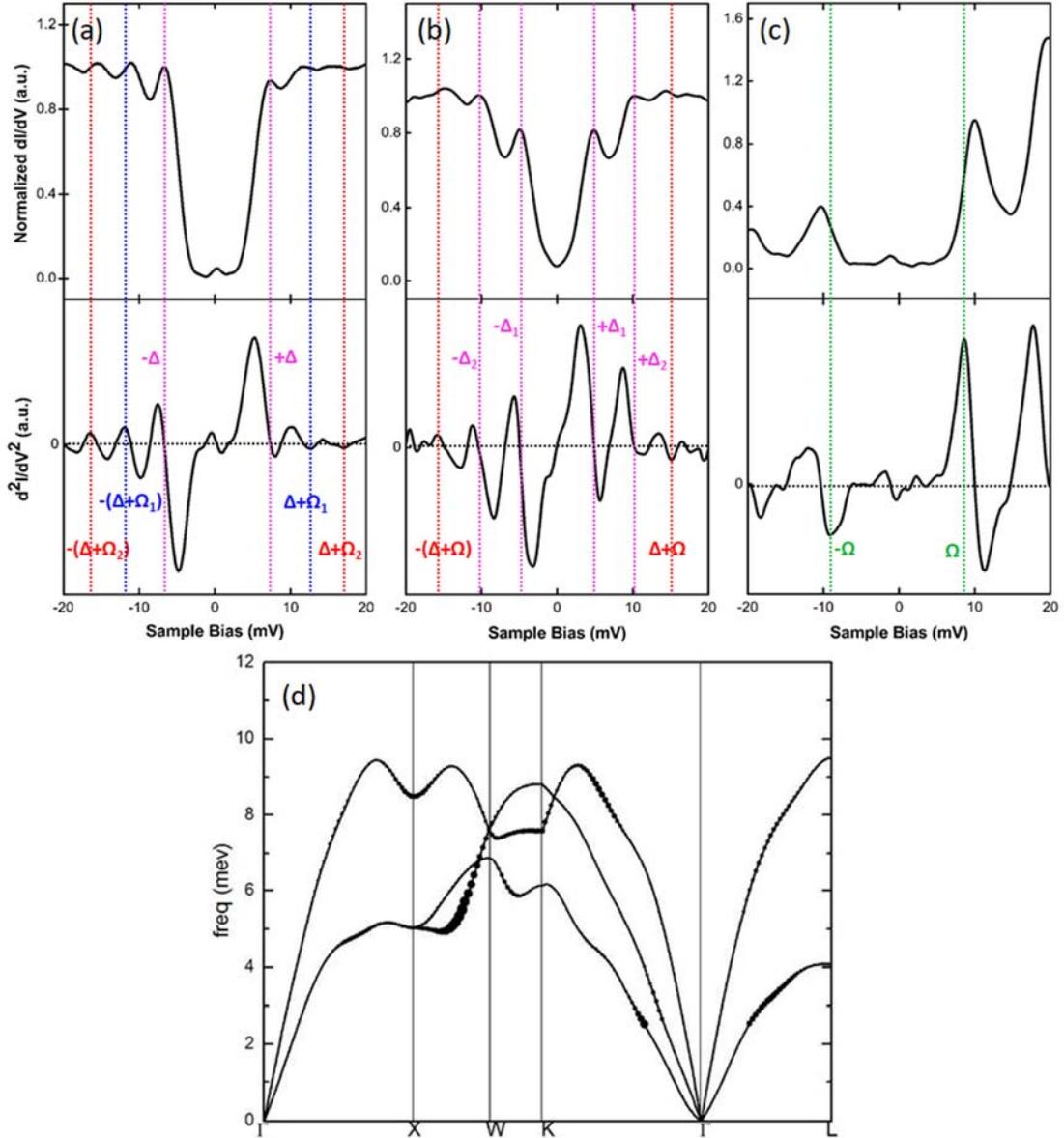

**Figure 4 | Characteristic phonon modes of the Pb islands.** (a)-(c) Representative normalized dI/dV and $d^2I/dV^2$ spectra in regions I, II and III, respectively. (a) The pink, blue, and red dashed lines show the approximate energy positions of $\pm\Delta$, $\pm(\Delta+\Omega_1)$, $\pm(\Delta+\Omega_2)$, respectively. $\Omega_1$ and $\Omega_2$ are the energies of phonon modes. (b) $\Delta_1$ and $\Delta_2$ are the energies of the two superconducting gaps. (c) The phonon excitation energies are marked by the green dash lines. (d) The phonon spectrum of bulk Pb, obtained by the first-principle density functional theory calculation. The size of dots is proportional to the strength of electron-phonon coupling, which exhibits a strong electron-phonon interaction between 4.5 meV and 7 meV.

## Characteristic phonon modes of the Pb islands

The characteristic energy scales of phonons (or other bosonic modes) that couple strongly with superconducting electrons can be derived from the second derivative of the tunneling current [11, 33, 34]. More specifically, if $\Omega$ is the energy of phonon, then a pair of a peak at -($\Delta+\Omega$) and a dip at +($\Delta+\Omega$) should develop in the $d^2I/dV^2$ spectrum, where $\Delta$ is the superconducting gap. The lower panels of Figs. 4(a)-(c) show the second derivative of tunneling current in three different regions of Pb islands. In region I, two pairs of dip-hump features are observed at ±($\Delta+\Omega$) ~ ±12.25meV and ±16.74meV, respectively. The corresponding phonon energies are found to be $\Omega_1$ = 5.28 meV and $\Omega_2$ = 9.76 meV, respectively. In region II, two similar phonon energy scales are observed at $\Omega_1$ = 5.30 meV and $\Omega_2$ = 10.58 meV. Changing from region I to region II, $\Omega_2$ is slightly enlarged, but $\Omega_1$ remains almost unchanged. In region III, the superconducting pairing is completely suppressed, a pair of gigantic phonon excitation peaks is observed at ±8.93 meV.

To determine the origin of these phonon modes, we calculated the phonon spectrum of bulk Pb. The phonon mode that couples most strongly with electrons, as shown in Fig. 4(d), is found to be between 4.5 meV to 7 meV, which is consistent with the energy scale $\Omega_1$. However, the energy of the second phonon mode $\Omega_2$ is above the full energy distribution of phonon modes in the bulk Pb, suggesting that this phonon mode may arise from the interface. In fact, according to the hyper-Raman scattering measurement [35, 36], $SrTiO_3$ has a TO1 phonon mode with a frequency about 10.8-10.9 meV. It has similar energy scale as $\Omega_2$ and the difference might be due to the finite size correction. Moreover, this higher energy phonon mode was not observed in the Pb islands grown on Si(111) [37, 38].

Our tunneling measurement reveals an enhanced superconductivity in Pb islands with a maximal superconducting gap about 11 meV. The superconducting energy gap in small superconducting Pb islands survives even above the maximal temperature (33 K) our measurement can be done. By fitting the superconducting energy gap with the BCS formula, we find that the transition temperature is higher than 40 K. This kind of $T_c$ enhancement has not been found in Pb films or islands grown on Si, GaAs, or other substrates [39], it indicates unambiguously that the SrTiO$_3$ substrate plays an important role.

Similar as for FeSe/SrTiO$_3$, two effects may have substantial contribution to the Tc enhancement in Pb/SrTiO$_3$. One is the charge transfer [6, 8, 10, 11, 40] from the substrate, the other is the strong enhancement of electron-phonon coupling [11, 12] induced by the interface or the strain caused by the lattice mismatch. The charge transfer may change the scattering potential of electrons near the interface, enabling the effective width of quantum well to be significantly larger than the thickness of Pb films. As already mentioned, this effective increase in the width of quantum well has indeed been observed in our measurement. The electron-phonon interaction is enhanced by the substrate because both T$_c$ and the ratio 2$\Delta$(0)/k$_B$T$_c$ in Pb/SrTiO$_3$ are significantly larger than in the bulk Pb samples.

**Conclusions**

To conclude, through the measurement of scanning tunneling spectroscopy, we found that the superconducting transition temperature of Pb islands grown on the SrTiO$_3$ (001) substrate is strongly enhanced in comparison with the bulk Pb. The superconducting energy gap increases with decreasing volume of Pb islands, but vanishes when the volume is below about 1200 nm$^3$. The largest energy gap of Pb island is around 10 meV, which is about 7 times larger than that of the bulk Pb superconductor. In the small Pb islands, the superconducting energy gap remains finite even at the highest temperature our measurement can be done (33 K). By fitting the temperature dependence of the energy gap with the BCS gap formula, we find that the superconducting transition temperature can be as high as 47 K, which is also about 7 times higher than that in the bulk Pb (7.2 K). However, it should be pointed out that from the tunneling measurement, we can only probe the density of states of low-energy excitations. In order to confirm that the gap states we observed have indeed the superconducting long-range order, transport (especially the resistivity) and thermodynamic (especially the magnetic susceptibility) measurements are desired.

**Methods**

**Sample synthesis.** The preparation of the substrate and the growth of Pb islands were carried out in a standard MBE system with a base pressure of $3\times10^{-10}$ Torr. The Nb-doped (001)-orientated single crystal SrTiO$_3$ substrate was heated to 1330K for 1 hour in UHV MBE chamber to obtain a clean and uniform surface. High-purity Pb source (99.99%) was deposited onto the SrTiO$_3$ substrate at 300K for 40 min with a rate of 0.5 ML/min from Knudsen cells. After deposition, the sample was annealed at 300K for 30 min to flatten the island top.

**STM/STS measurements.** For the STM/S measurements, a low-temperature STM system (Unisoku USM-1300), which is capable of cooling the sample down to 400mK and applying a magnetic field up to 13 T perpendicular to the sample surface, was used for the topographic observation and the scanning tunneling spectroscopy (STS) measurement. All STM topographic images were taken at a constant current of 20pA. The low bias (energy) dI/dV tunneling spectra were measured using a lock-in technique with a bias modulation of 0.5 mV at 879.321 Hz. The large-energy scale dI/dV tunneling spectra were measured using a bias modulation of 5 mV. A tungsten tip, whose oxide layer covering the apex was removed in situ by electron-beam heating, was used in all topographic observations and STS measurements.

## Acknowledgments

This work was conducted at the analysis and test center and the Department of Physics of Huazhong University of Science and Technology in China. HY and MHP acknowledge the financial supported by National Natural Science Foundation of China (Project Code 11574095 and 11604106). TX acknowledge the support by the National R&D Program of China (2017YFA0302900) and the National Natural Science Foundation of China (11474331).

Support Information for

# Scanning tunneling microscopic evidence of interface enhanced high-T$_c$ superconductivity in Pb islands grown on SrTiO$_3$

Haigen Sun[1], Zhibin Shao[1], Zongyuan Zhang[1], Yan Cao[1], Shaojian Li[1], Xin Zhang[1], Qi Bian[1], Habakubaho Gedeon[1], Lijuan Liu, Hui Yuan[1], Jianfeng Zhang[2], Kai Liu[2], Zhong-Yi Lu[2], Tao Xiang[3,4], Qi-Kun Xue[4,5] and Minghu Pan[1]

[1]School of Physics, Huazhong University of Science and Technology, Wuhan 430074, China.

[2]Department of Physics，Renmin University of China，Beijing 100872，China

[3]Institute of Physics, Chinese Academy of Sciences, Beijing 100190, China

[4]Collaborative Innovation Center of Quantum Matter, Beijing 100084, China

[5] Department of Physics, Tsinghua University, Beijing 100084, China

Haigen Sun, Zhibin Shao, Zongyuan Zhang contributed equally to this work.
Corresponding author. E-mail: qkxue@mail.tsinghua.edu.cn (Q.-K. X.); minghupan@hust.edu.cn (M. H. P.).

## Methods

**Sample synthesis.** The preparation of the substrate and the growth of Pb islands were carried out in a standard MBE system with a base pressure of 3x10$^{-10}$ Torr. The Nb-doped (001)-orientated single crystal SrTiO$_3$ substrate was heated to 1330K for 1 hour in UHV MBE chamber to obtain a clean and uniform surface. High-purity Pb source (99.99%) was deposited onto the SrTiO$_3$ substrate at 300K for 40 min with a rate of 0.5 ML/min from Knudsen cells. After deposition, the sample was annealed at 300K for 30 min to flatten the island top.

**STM/STS measurements.** For the STM/S measurements, a low-temperature STM system (Unisoku USM-1300), which is capable of cooling the sample down to 400mK and applying a magnetic field up to 13 T perpendicular to the sample surface, was used for the topographic observation and the scanning tunneling spectroscopy (STS) measurement. All STM topographic images were taken at a constant current of 20pA. The low bias (energy) dI/dV tunneling spectra were measured using a lock-in technique with a bias modulation of 0.5 mV at 879.321 Hz. The large-energy scale dI/dV tunneling spectra were measured using a bias modulation of 5 mV. A tungsten tip, whose oxide layer covering the apex was removed in situ by electron-beam heating, was used in all topographic observations and STS measurements.

**1, dI/dV curves and the fitting by the Dynes function**

In order to determine the gap value quantitatively, we fit the tunneling spectra using the Dynes function:

$$\frac{dI}{dV} \propto \frac{1}{kT} \int_{-\infty}^{\infty} dE \ \text{Re}\left[\frac{|E - i\Gamma|}{\sqrt{(E - i\Gamma)^2 - \Delta^2}}\right] \cosh^{-2}\frac{E + eV}{2kT}$$

where $\Gamma$ is an effective broadening parameter. Figure S1 shows, as an example, the fitting curves for the dI/dV data measured on a Pb island in Region I.

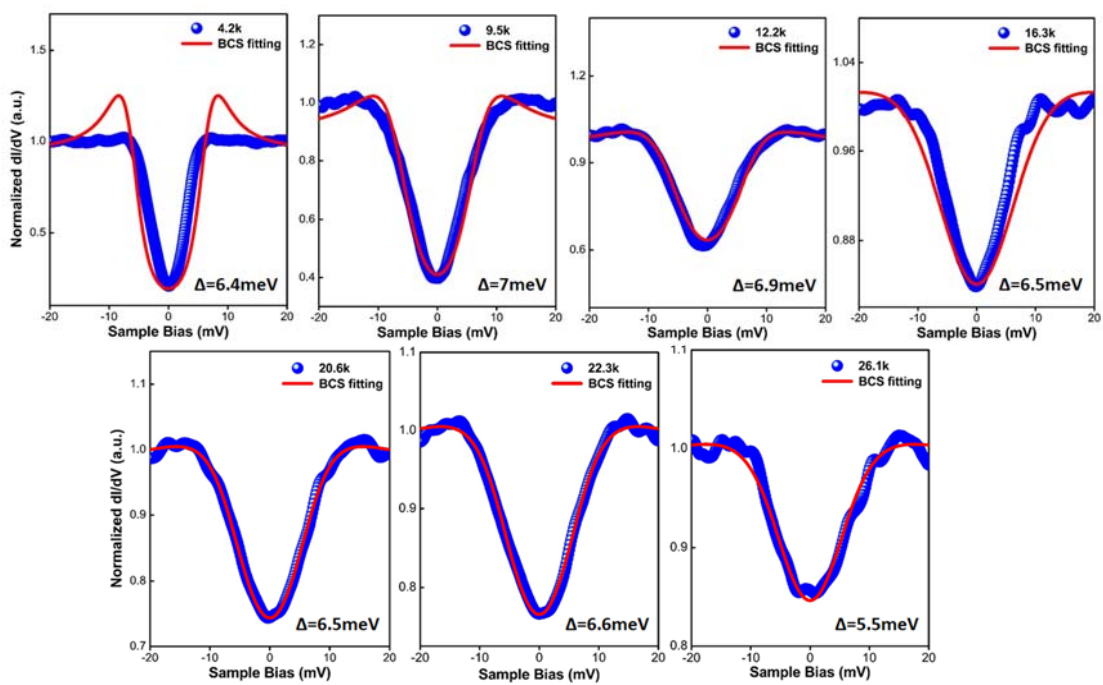

**Figure S1** | The measurement data of dI/dV (blue dots) taken at a Pb island with the volume V=4628 nm³ and the fitting curves obtained using the Dynes function. The gap parameters obtained from the fitting are shown inside each figure.

## 2, Spectra for Pb islands with different volumes

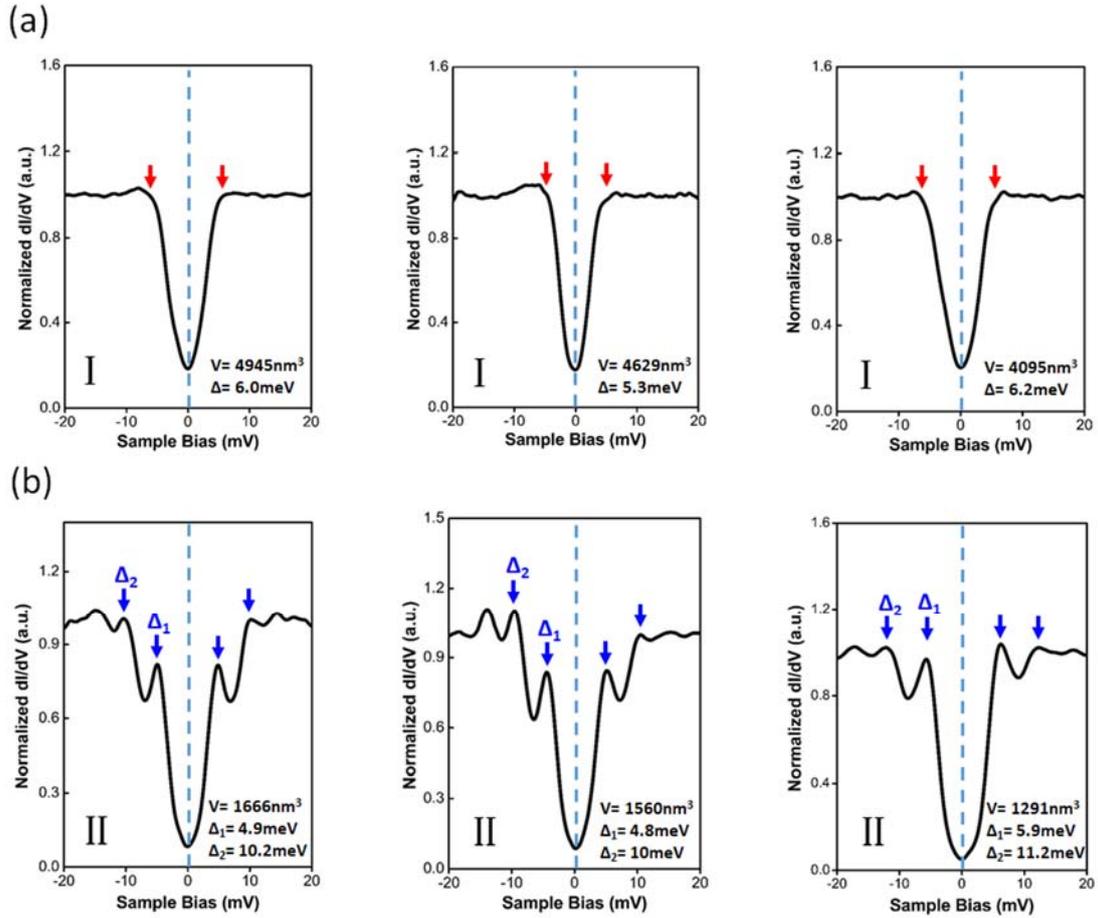

**Figure S2** │ Variance of the differential conductance spectra with the volume of Pb islands. (a)-(b) The spectra taken at the centers of Pb islands in two different regions as depicted in Fig. 3(a). In region I, the spectra exhibit a tiny coherent peak at either positive or negative bias side. In region II, the spectra show two supconducting peaks as marked by blue arrows. The gap values $\Delta$ are determined by the energy differences between the left gap edges marked by the arrows and the blue dashed lines.

## 3, dI/dV spectra taken at different locations on the Pb islands

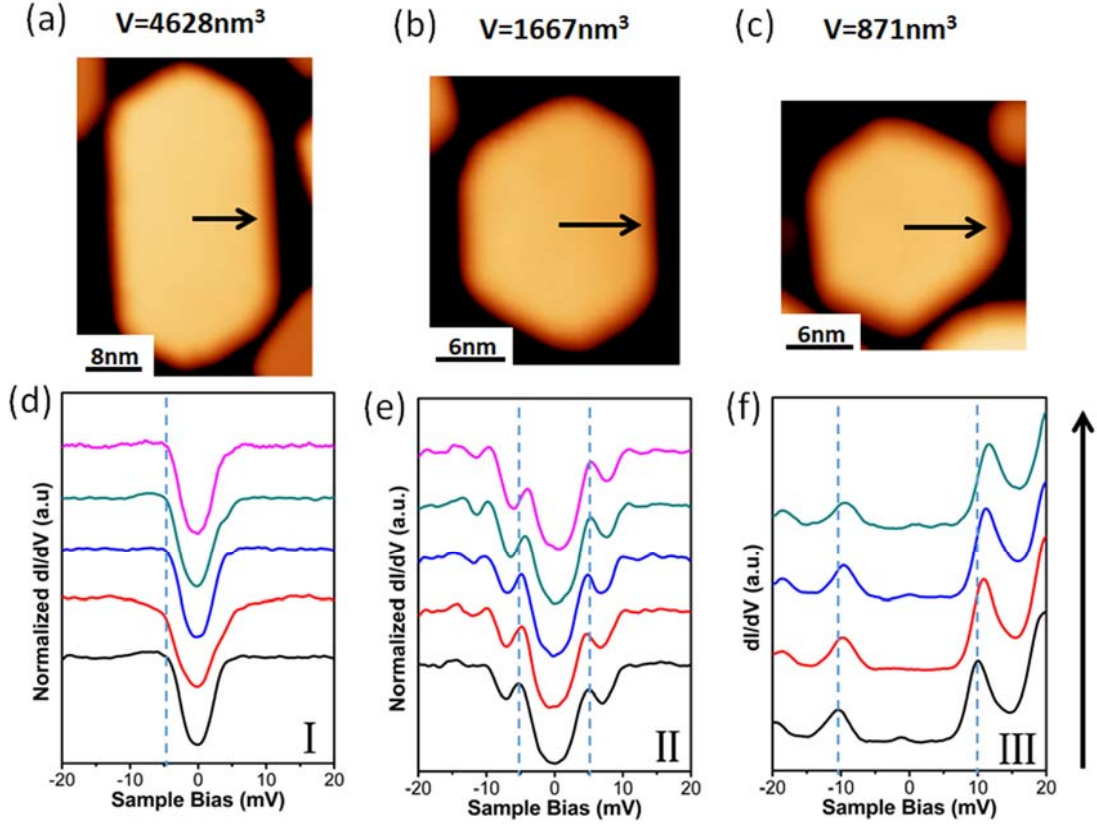

**Figure S3** │ (a)-(c) Topographic images of three representative Pb islands with different volumes. (d)-(f) The normalized dI/dV spectra measured along the black arrow marked in (a)-(c). The vertical dash lines are provided to guide the eye. In contrast to the normalized dI/dV spectra shown in (d), the conductance peaks in (e) taken at the sample edge is slightly suppressed and shifted inwards in comparison with that at the center.

## 4, Temperature dependence of the spectra in region II

The islands in Region II exhibit a two-gap structure. The smaller gap becomes invisible above 9.5 K or slightly higher temperature, and the larger gap is gradually suppressed with increasing temperature. Due to the limitation of our STM instrument, our measurement cannot exceed 33 K. At 33 K, the gap feature still exists. We fit the tunneling spectra by utilizing a generalized Dynes function with two isotropic s-wave gaps

$$G = x\frac{dI_1}{dV} + (1-x)\frac{dI_2}{dV}$$

$$\frac{dI_{1(2)}}{dV} \propto \int_{-\infty}^{\infty} dE \; \text{Re}\left[\frac{|E - i\Gamma_{1(2)}|}{\sqrt{(E - i\Gamma_{1(2)})^2 - \Delta_{1(2)}^2}}\right] \cosh^{-2}\frac{E + eV}{2kT}$$

where G is the differential conductivity. $I_1$ and $I_2$ are the tunneling current contributed by the larger and smaller gaps, respectively. x is the relative contribution of the smaller gap. Fig. S4 shows, as an example, the measurement data with the fitting curves for a Pb island in Region II.

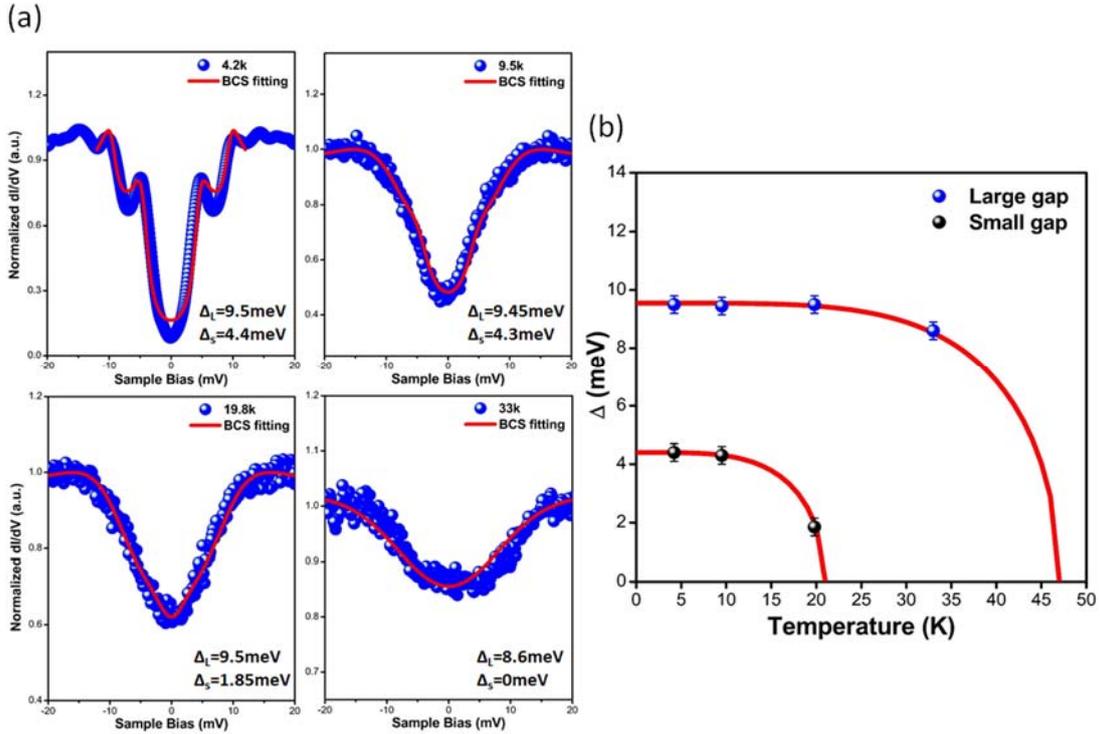

**Figure S4** | (a) The spectra measured at four different temperatures. The blue filled cycles represent the experimental data while the BCS fittings are shown by the red curves. The gap parameters obtained from the fitting are shown in each figure. (b) Temperature dependence of the two gaps obtained by fitting the experimental data using Dynes function. By fitting the data with the BCS gap function (red curve), $T_c$ for the two gaps are found to be about 21K and 47 K respectively.